\documentclass[notitlepage,amsfonts,showpacs,tightenlines,aps,12pt,floatfix]
{revtex4}
\usepackage{bm}
\usepackage{epsfig}
\usepackage{graphicx}
\begin{document}
\title{Duality Violation and the
$K\rightarrow\pi\pi$ Electroweak Penguin Operator Matrix Elements
from Hadronic $\tau$ Decays}

\author{Kim Maltman$^{a,b}$, D. Boito$^{c,d}$, O. Cata$^e$, M. Golterman$^f$, 
M. Jamin$^{d,g}$, J. Osborne$^f$, S. Peris$^{c,f}$}

\affiliation{${^a}$Math \& Stats, York U., Toronto, CANADA\\
$^b$CSSM, U. Adelaide, Adelaide, Australia\\
${^c}$Dept. Physics , U. Aut\`onoma de Barcelona, Barcelona, Spain\\
$^d$IFAE, U. Aut\`onoma de Barcelona, Barcelona, Spain\\
${^e}$Dept. de Fisica Te\'orica, IFIC, U. Valencia-CSIC, Valencia, Spain\\
${^f}$Dept Physics and Astronomy, SFSU, San Francisco, USA\\
${^g}$ICREA U. Aut\`onoma de Barcelona, Barcelona, Spain}

\begin{abstract}
We discuss a preliminary study of the impact
of duality violations on extractions from
$\tau$ decay data of the $D=6$ VEVs which
determine chiral limit Standard Model
$K\rightarrow\pi\pi$ matrix elements of the
electroweak penguin operators.
\end{abstract}


\maketitle


\section{Introduction}
In the Standard Model (SM), 
$\epsilon^\prime/\epsilon$ is dominated by contributions from the 
gluonic and electroweak penguin (EWP) operators, $Q_6$ and
$Q_8$. 
In the $SU(3)$ chiral limit, the $K\rightarrow\pi\pi$ matrix elements of 
the EWP operators $Q_{7,8}$, are determined
by two 4-quark VEVs, $\langle {\cal O}_1\rangle$
and $\langle {\cal O}_8\rangle$, which also 
determine the dimension $D=6$ part of the OPE of
the flavor $ud$ V-A correlator difference 
$\Delta\Pi\equiv \Pi^{(0+1)}_V-\Pi^{(0+1)}_A$~\cite{dgbasic},
where the superscript $0+1$ denotes
the sum of spin $J=0$ and $1$ components and, 
with $J_{V/A}^\mu$ the standard V or A $ud$ current,
the scalar correlators $\Pi^{(J)}_{V/A}$ are defined via
\begin{equation}
i \int d^4x \, e^{iq\cdot x} 
\langle 0|T(J_{V/A}^{\mu}(x) J_{V/A}^{\nu}(0)^\dagger)|0\rangle \, 
\equiv\,  
(q^\mu q^\nu - g^{\mu\nu} q^2 ) \, \Pi^{(1)}_{V/A}(q^2)
+ q^{\mu} q^{\nu} \, \Pi^{(0)}_{V/A}(q^2)\ .\label{twoptdefn}
\end{equation}
Since $\left[\Delta\Pi\right]^{OPE}_{D=6}$ is
strongly dominated by the contribution involving
$\langle {\cal O}_8\rangle$, which VEV also dominates the
chiral limit $Q_8$ matrix element, the extraction of 
$\left[\Delta\Pi\right]_{D=6}^{OPE}$ is of considerable
phenomenological interest, and a number of 
dispersive and finite energy sum rule (FESR) analyses
have attempted it~\cite{cdgm01,vmac6c8etc,cgm03,ppg10}.
$\tau$ decay data plays a key role in these analyses 
since the spectral function of $\Delta\Pi$, 
$\Delta\rho (s)\, =\, {\frac{1}{\pi}}\, Im\, \Delta\Pi (s+i\epsilon)$,
is directly measurable for $s\leq m_\tau^2$ 
in non-strange hadronic $\tau$ decays. 
Explicitly, in the SM, with $S_{EW}$ a short-distance EW correction,
$y \equiv s/m_\tau^2$, $w_T(y)=(1-y)^2(1+2y)$ and
$R^{ud}_{V/A} \equiv \Gamma [\tau^- \rightarrow \nu_\tau
\, {\rm hadrons}^{ud}_{V/A}]/ \Gamma [\tau^- \rightarrow
\nu_\tau e^- {\bar \nu}_e ]$, one has, for the continuum 
(non-$\pi$-pole) part of $\Delta\rho$~\cite{tsai}
\begin{equation}
\Delta\rho (s)\, =\, {\frac{m_\tau^2}
{12\pi^2\vert V_{ud}\vert^2 S_{EW} w_T(y)}}
\,  {\frac{dR^{ud}_{V-A}}{ds}}\ .
\label{rhofromdrds}\end{equation}

Dispersive analyses employ the unsubtracted dispersion
relation for $\Delta\Pi$ and require either 
assumptions about the saturation of the dispersion integral 
within the range kinematically 
accessible in $\tau$ decays, or supplementary
constraints on $\Delta\rho (s)$ for $s>m_\tau^2$, such as those provided by the
Weinberg sum rules~\cite{weinbergsr} and the DGMLY $\pi$ electromagnetic (EM)
self-energy sum rule~\cite{dgmlysr} (see, e.g., Ref.~\cite{cdgm01}
for details). Higher dimension ($D>6$) contributions to
$\Delta\Pi (Q^2)$ must also be considered. 
These problems are avoided in the FESR approach, which relies
on $\Delta\Pi (s)$ having no 
kinematic singularities and hence satisfying
the FESR relation
\begin{equation}
\int\, ds\, w(s)\, \Delta\rho (s)\ =\, {\frac{-1}{2\pi i}}\,
\oint_{\vert s\vert =s_0}ds\, w(s)\, \Delta\Pi (s)\ ,
\label{fesrreln}\end{equation}
for any $s_0$ and any $w(s)$ analytic in the region of the contour. 
For sufficiently large $s_0$, the OPE should become reliable on the RHS.
Choosing polynomial weights $w(s)$ with degree $N$ strongly suppresses 
OPE contributions with $D>2N+2$. For sub-asymptotic $s_0$, OPE breakdown, 
or duality violation (DV) is expected. In fact, even for
$s_0\sim m_\tau^2$, sizeable $s_0$-dependent deviations between the 
LHS and OPE versions of the RHS are found for the $w(s)=1$
V and A analogues of Eq.~\ref{fesrreln}~\cite{kmfesr,ouralphasprelim10}. 
These are strongly suppressed for analogues employing pinched weights 
($w(s)$ with a zero at $s=s_0$)~\cite{kmfesr}, indicating that at
scales $\sim m_\tau^2$ DVs are localized to the vicinity of the 
timelike axis. With this in mind, the analysis
of Ref.~\cite{cgm03} (CGM) employed doubly pinched weights, 
checking the $s_0$-dependence of the match between the weighted 
spectral integrals and optimized OPE fit as a test of the 
self-consistency of the assumed neglect of residual DV contributions.
Figure~1 shows the resulting residuals, 
$\left[I^w_{OPE}(s_0)-I^w_{OPAL}(s_0)\right]/\delta I^w_{OPAL}(s_0)$,
over an expanded $s_0$ range, for the
two weights, $w_1$ and $w_2$ of the ``maximally safe'' CGM 
analysis based on OPAL data~\cite{opaldata}. (We focus
here on OPAL data due to a problem with the ALEPH covariance 
matrices~\cite{ouralphasprelim10} which is the subject of 
ongoing reanalysis.) 
$I^w_{OPAL,OPE}(s_0)$ are the LHS and RHSs of Eq.~\ref{fesrreln}
and $\delta I^w_{OPAL}(s_0)$ the uncertainty on $I^w_{OPAL}(s_0)$.
It is obvious that residual DVs, though not evident {\it within errors} 
above $s_0\sim 2\ {\rm GeV}^2$, become non-negligible 
below this point. Small residual DV contibutions are thus
expected in the $s_0> 2\ {\rm GeV}^2$ CGM fit window as well.
Lacking a model for DVs, analyses such as CGM were unable
to estimate the systematic uncertainty associated
with neglecting these contributions.

\begin{figure}\label{fig1}
\caption{Residuals for the CGM ``maximally safe'' fit, with no DVs}
\begin{minipage}[t]{0.48\linewidth}
\rotatebox{270}{\mbox{
  \includegraphics[width=2in]{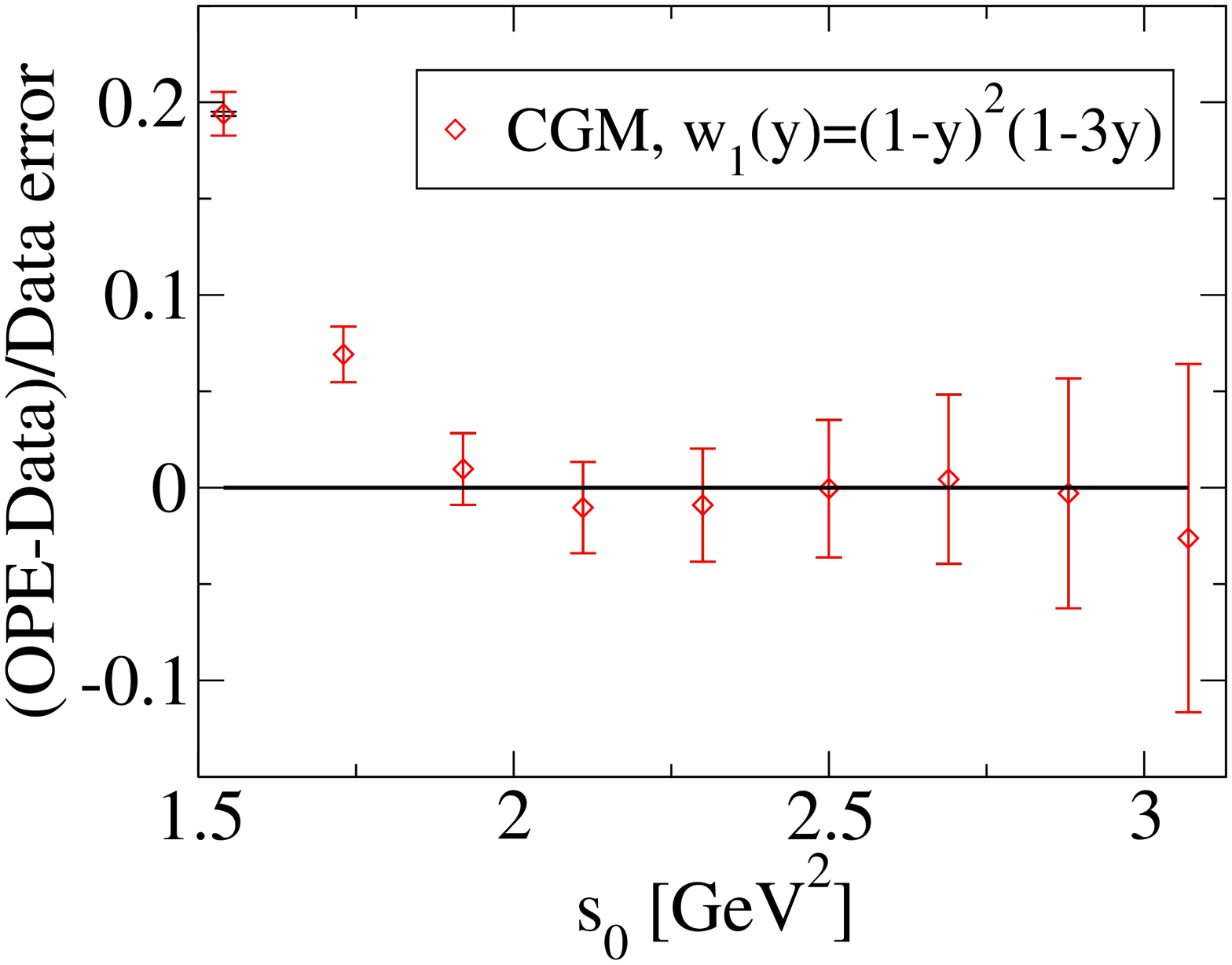}
}}
\end{minipage}
\hfill
\begin{minipage}[t]{0.48\linewidth}
\rotatebox{270}{\mbox{
  \includegraphics[width=2.in]{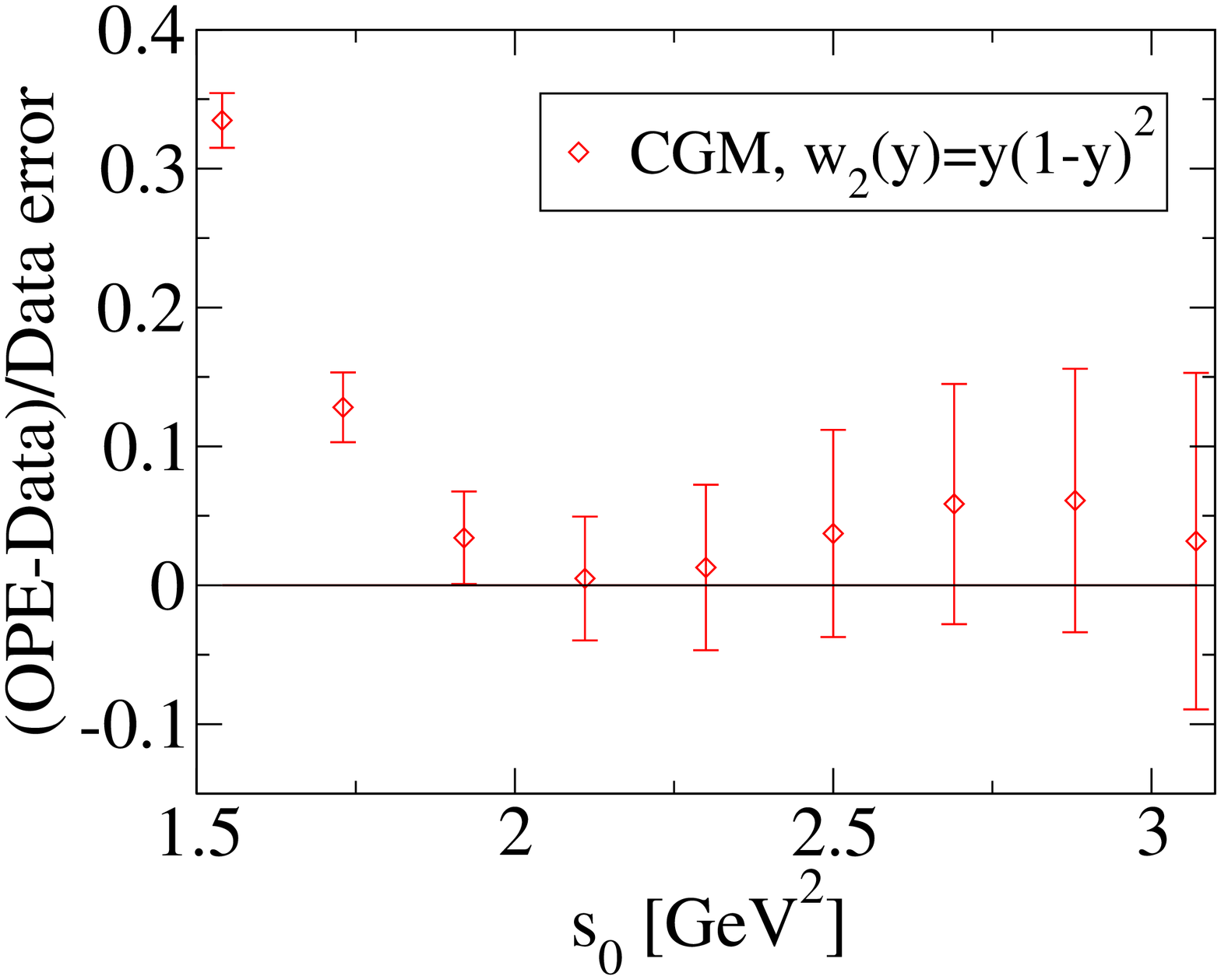}
}}
\end{minipage}
\end{figure}

\section{Incorporating Duality Violations}
In Refs.~\cite{cgp08}, a model for DV spectral contributions was developed.
The model builds on earlier work in Refs.~\cite{basicdvmodelideas} and
is motivated by large-$N_c$ and Regge-based resonance spacing ideas.
The model leads to ans\"atze $\rho_{V/A}(s)=\rho^{(D=0)}_{OPE}(s)
+\rho_{V/A}^{DV}(s)$, $s>s_{min}$, for the V and A spectral functions, where
the DV contributions have the form
\begin{equation}
\rho^{DV}_{V/A}(s)=\kappa_{V/A} e^{-\gamma_{V/A} s} \sin (
\alpha_{V/A}+\beta_{V/A}s)\ .
\label{dvmodelform}\end{equation}
In Refs.~\cite{ppg10} the impact of DVs on previous V-A analyses
was investigated using a {\it single} DV ansatz of the form 
Eq.~\ref{dvmodelform} for the V-A difference $\Delta\rho (s)$.
This involves the implicit additional assumption
that $\beta_V\simeq\beta_A$ and $\gamma_V\simeq\gamma_A$,
allowing the $8$-parameter V-A difference to be re-written in the
effective $4$-parameter form, Eq.~\ref{dvmodelform}.
We avoid this additional assumption
and fit the V and A DV parameter sets separately, as part of
a combined V, A fit which also determines
the OPE parameters $\alpha_s$, $\langle \alpha_s G^2\rangle$,
and the relevant $D=6$ and $8$ V and A channel effective condensates.
We find central DV parameter fit values not in
good accord with the expectations $\beta_V\simeq\beta_A$, 
$\gamma_V\simeq\gamma_A$.

Our analysis employs $w(s)$ up to degree $3$,
including $w(s)=1$, which is optimally
sensitive to the DV contributions. The resulting fits provide
excellent matches between the OPAL spectral integrals and 
optimized OPE+DV fit forms for all $w(s)$ employed and
all $s_0$ down to a fit window minimum $s_0^{min}\sim 1.4-1.5\ {\rm GeV}^2$.
Though so far aimed at extracting $\alpha_s$, and not optimized for extracting
$D=6,8$ V-A condensates, the analysis nonetheless provides preliminary 
results for these
quantities. Since the fits provide a prediction for
$\Delta\rho (s)$ for $s>s_0^{min}$,
and hence also above $s=m_\tau^2$, we can test our results against 
the Weinberg and DGMLY sum rules, which constraints have {\it not}
been incorporated in performing the fits. The first and second
Weinberg sum rules are written in a form with RHSs equal to
zero; for the RHS of the DGMLY sum rule we employ
the $SU(2)$ chiral limit value $-8\pi f_\pi^2\left[\delta m_\pi^2\right]_{EM}
/3\alpha_{EM}\, =\, -0.0109(15)\ {\rm GeV}^4$~\cite{ppg10}.
The results of these tests are shown in Figure~2,
with $s_0$ the point beyond
which the fitted form of $\Delta\rho (s)$ is employed in the
relevant spectral integral. Below this point, experimental
data are used. The dotted and solid black lines in the
third panel show the central DGMLY sum rule RHS and error.
All three sum rules should be satisfied
for all $s_0$ in our fit window. This is evidently the
case, giving us good confidence in the results for the fitted OPE
parameters as well.

\begin{figure}\label{fig2}
\caption{The Weinberg and DGMLY sum rule tests, the DGMLY case
involving the weight $w(s)=s\, log(s/\Lambda^2)$.}
\begin{minipage}[t]{0.26\linewidth}
\rotatebox{270}{\mbox{
  \includegraphics[width=1.7in]{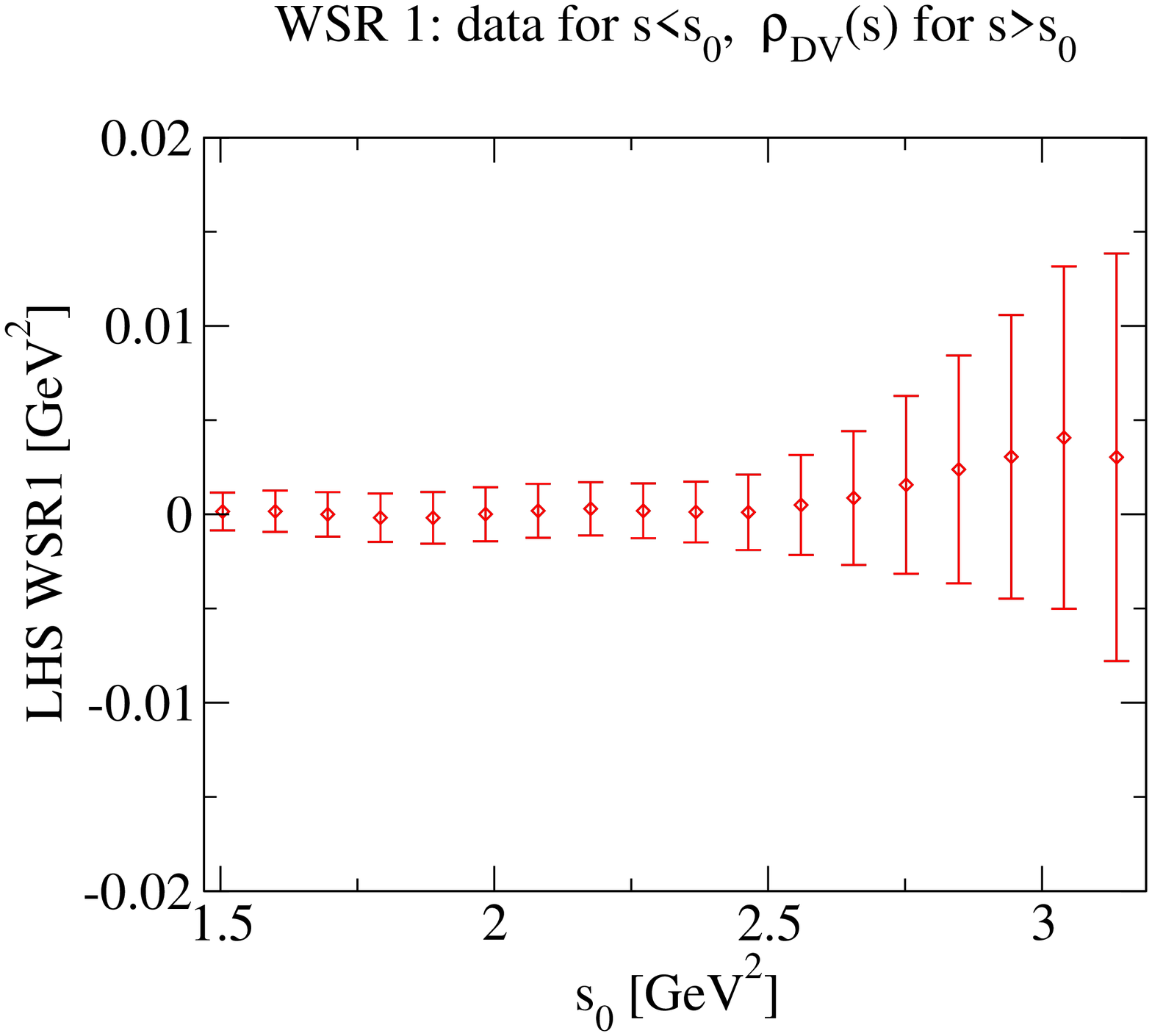}
}}
\end{minipage}
\ \ \ \ \hfill
\begin{minipage}[t]{0.26\linewidth}
\rotatebox{270}{\mbox{
  \includegraphics[width=1.7in]{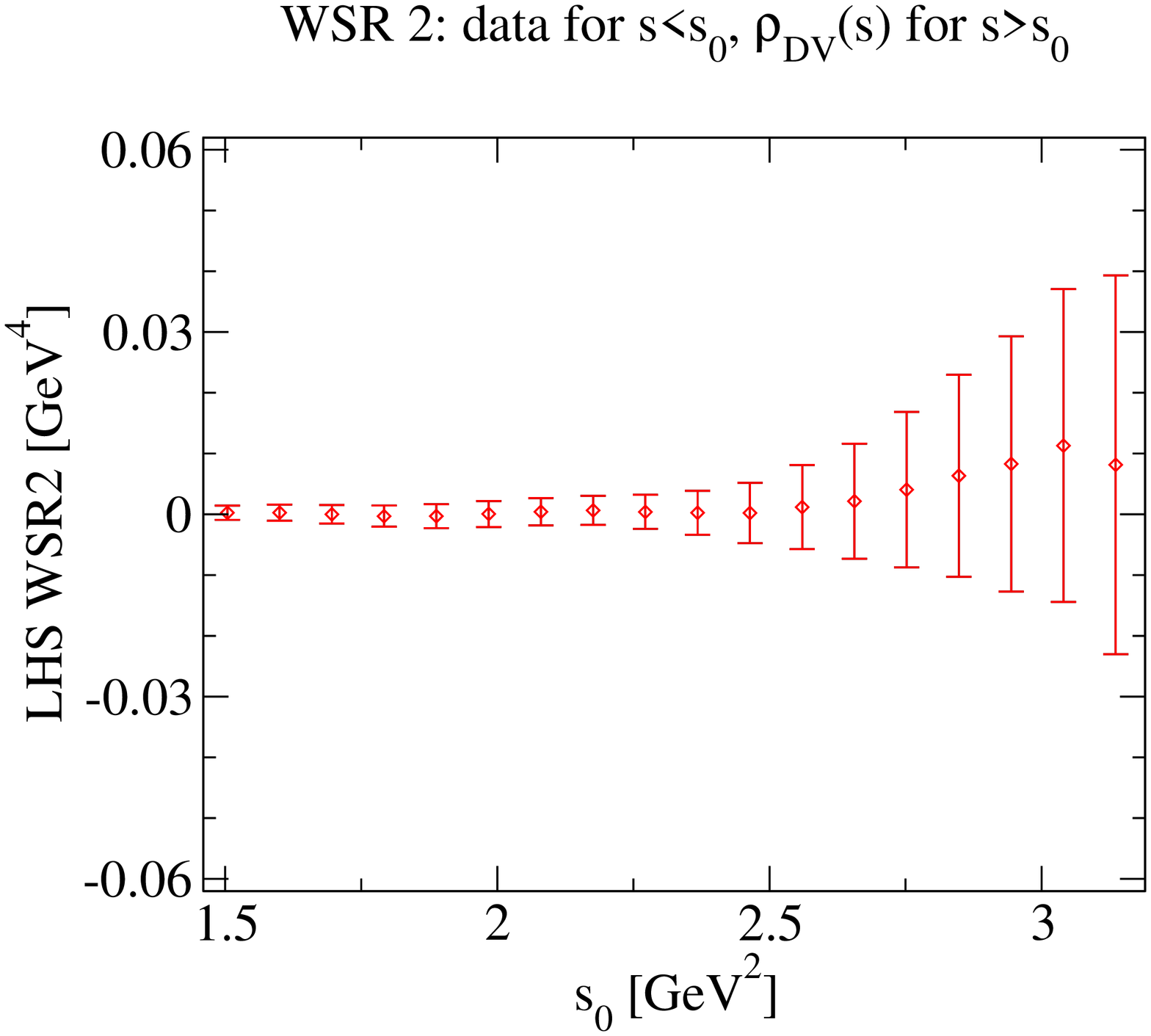}
}}
\end{minipage}
\ \ \ \ \hfill
\begin{minipage}[t]{0.26\linewidth}
\rotatebox{270}{\mbox{
  \includegraphics[width=1.7in]{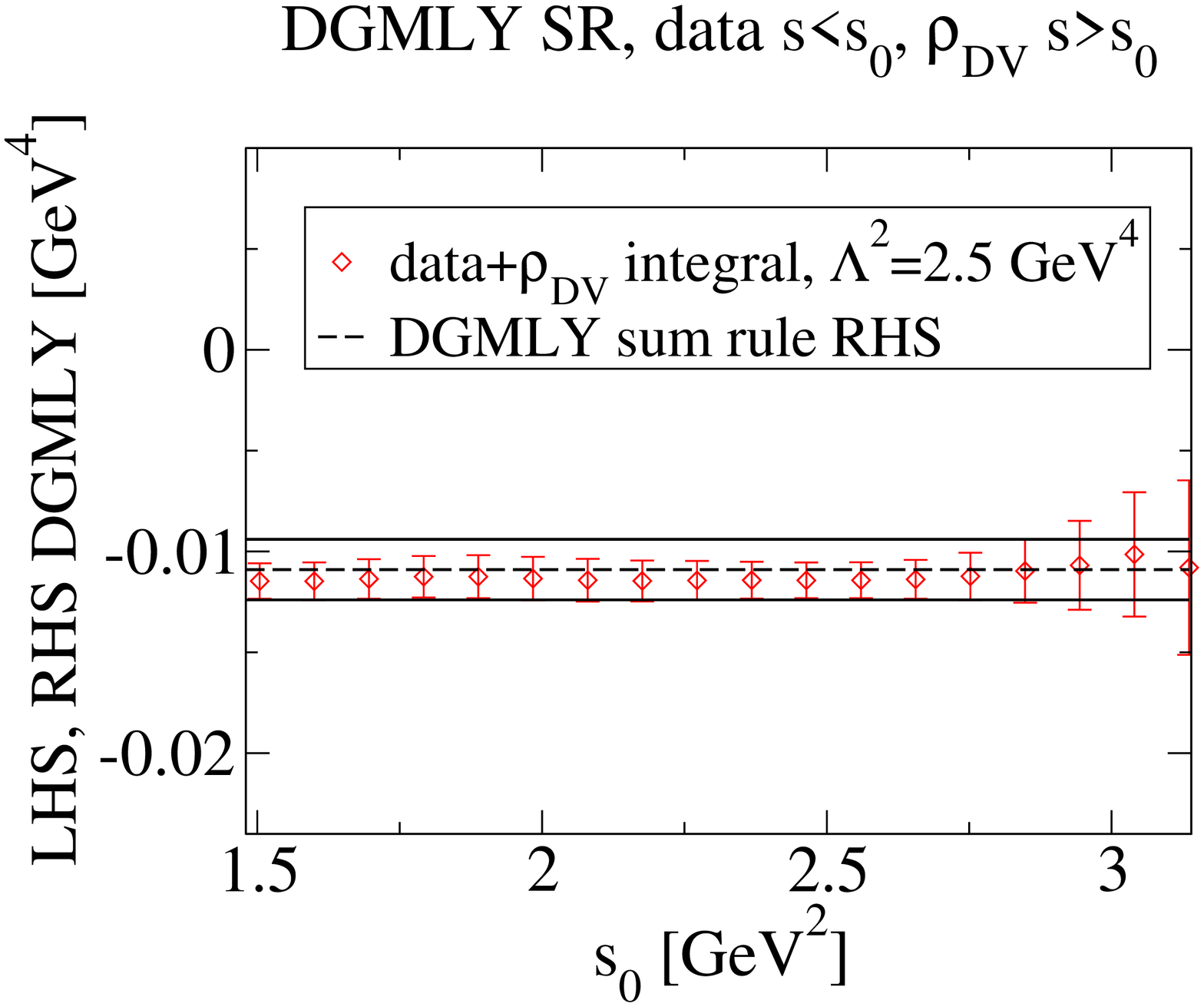}
}}
\end{minipage}
\end{figure}

As an illustration of our preliminary results, we quote the effective
$D=6$ V-A condensate $C_6^{V-A}$ which results from an update
of the CGM OPAL fit incorporating DVs and fitting
$\rho_V$, $\rho_A$, and the $w(y)=1$, $1-y$ and
$(1-y)^2$ ($y=s/s_0$) FESRs with the CIPT scheme for the 
truncated $D=0$ OPE series and $s_0^{min}=1.504\ {\rm GeV}^2$. 
$C_6^{V-A}$ is defined by 
$\left[\Delta\Pi (Q^2)\right]^{OPE}_{D=6}=C_6^{V-A}/Q^6$.
We find $C_6^{V-A}=-0.0057(18)\ {\rm GeV}^6$, c.f. 
the CGM no-DVs maximally safe analysis OPAL-based value
$-0.0054(8)\ {\rm GeV}^6$. Values somewhat
larger in magnitude, with larger errors, are obtained from
analyses excluding $w(y)=1-y$, including those employing
FOPT for the $D=0$ series. We emphasize that these
results are preliminary, and that a dedicated V-A analysis, aimed
at reducing the errors, is in progress. Readers noting
the $20-25\%$ difference between the values quoted above and those
obtained from the ALEPH-based analyses of CGM (neglecting DVs)
and Ref.~\cite{ppg10} (including DVs approximately) should bear
in mind that the ALEPH and OPAL $4\pi$ data differ significantly 
in the upper part of the spectrum, with the OPAL data agreeing better with
expectations based on CVC and recent preliminary BaBar and SND
$4\pi$ electroproduction data~\cite{epem4pi}.

\acknowledgments
This work was supported 
by the Natural Sciences and
Engineering Research Council of Canada, the US DOE,
MICINN (Spain) under Grant FPA 2007-60323, the
Spanish Consolider Ingenio 2010 Program CPAN (CSD2007-00042),
CICYTFEDER-FPA2008-01430, SGR2005-00916 and the Programa de
Movilidad PR2010-0284.


\begin{thebibliography}{99}

\bibitem{dgbasic}
J.F. Donoghue and E. Golowich, \emph{Phys. Lett.}
\textbf{478}, 172 (2000).

\bibitem{cdgm01}V. Cirigliano {\it et al.}, 
\emph{Phys. Lett.}
\textbf{B522}, 245 (2001). 

\bibitem{vmac6c8etc}S. Peris, B. Phily, E. de Rafael, 
\emph{Phys. Rev. Lett.} \textbf{86}, 14 (2001); J. Bijnens, E. Gamiz,  
J. Prades, \emph{JHEP} \textbf{0110}, 009 (2001); M. Knecht, S. Peris, 
E. de Rafael, \emph{Phys. Lett.} \textbf{B508}, 117 (2001); B.L. Ioffe,  
K.N. Zyablyuk, \emph{Nucl. Phys.} \textbf{A687}, 437 (2001); K. N.
Zyablyuk, \emph{Eur. Phys. J.} \textbf{C38}, 215 (2004); J. Roja, 
J.I. Latorre, \emph{JHEP} \textbf{0401}, 055 (2004); C. A. Dominguez, 
K. Schilcher, \emph{Phys. Lett.} \textbf{B581}, 193 (2004); S. Friot,
D. Greynat, E. de Rafael, \emph{JHEP} \textbf{0410}, 043 (2004);
S. Narison, \emph{Phys. Lett.} \textbf{B624}, 223 (2005); S. Schael
{\it et al.} (ALEPH), \emph{Phys. Rep.} \textbf{421},
191 (2005); J. Bordes {\it et al.}, \emph{JHEP} \textbf{0602}, 037 (2006);
P. Masjuan, S. Peris, \emph{JHEP} \textbf{0705}, 040 (2007);
A.A. Almasy, K. Schilcher, H. Spiesberger, \emph{Eur. Phys. J.}
\textbf{C55}, 237 (2008).

\bibitem{cgm03}V. Cirigliano {\it et al.}, \emph{Phys. Lett.}
\textbf{B555}, 71 (2003); V. Cirigliano, E. Golowich and K. Maltman,
\emph{Phys. Rev.} \textbf{D68}, 054013 (2003).

\bibitem{ppg10}M. Gonzalez-Alonso, 
A. Pich and J. Prades, \emph{Phys. Rev.}
\textbf{D81}, 074007 (2010); \emph{Phys. Rev.} \textbf{D82}, 014019 (2010).

\bibitem{tsai}Y. Tsai, \emph{Phys. Rev.} \textbf{D4}, 2821 (1971).

\bibitem{weinbergsr}S. Weinberg, 
\emph{Phys. Rev. Lett.} \textbf{18},
507 (1967).

\bibitem{dgmlysr}T. Das {\it et al.}, 
\emph{Phys. Rev. Lett.} \textbf{18},
759 (1967).

\bibitem{kmfesr}K. Maltman, \emph{Phys. Lett.} \textbf{B440}, 
367 (1998);
C. A. Dominguez and K. Schilcher, \emph{Phys. Lett.} \textbf{448},
93 (1999).

\bibitem{ouralphasprelim10}D. Boito {\it et al.}, 
arXiv: 1011.4426 [hep-ph];
arXiv: 1103.4194 [hep-ph]

\bibitem[Opal (1999)]{opaldata}K. Ackerstaff {\it et al.} (OPAL), 
\emph{Eur. Phys. J.}
\textbf{C7}, 571 (1999). 

\bibitem{cgp08}O. Cat\`a, M. Golterman and S. Peris, 
\emph{JHEP} \textbf{0508}, 076 (2005); \emph{Phys. Rev.} \textbf{D77}, 
093006 (2008); \emph{Phys. Rev.} \textbf{D79}, 053002 (2009).

\bibitem{basicdvmodelideas}B. Blok, M.A. Shifman and D.X. Zhang, 
\emph{Phys. Rev.} \textbf{D57}, 2691 (1998) [erratum: {\it ibid.}
19901 (1999)], I.I.Y. Bigi, {\it et al.}, \emph{Phys. Rev.}
\textbf{D59}, 054011 (1999).

\bibitem{epem4pi}See, e.g., V.P. Druzhinin {\it et al.}, 
arXiv:1105.4975
and the poster session presentation by L. Kardapoltsev at PANIC 2011.

\end{thebibliography}
\end{document}